\documentstyle[aps,epsf,twocolumn]{revtex}

\newcommand{\be}{\begin{eqnarray}}
\newcommand{\ee}{\end{eqnarray}}

\begin{document}

\draft
\title{\bf  BRST Quantization, Strong CP Violation, the U(1) Problem  and
$\theta$ Vacua}

\author{{\bf Hidenaga Yamagishi}$^1$
and {\bf Ismail Zahed}$^2$}

\address{$^1$4 Chome 11-16-502, Shimomeguro, Meguro, Tokyo, Japan. 153;\\
$^2$Department of Physics, SUNY, Stony Brook, New York 11794, USA.}
\date{\today}
\maketitle

\begin{abstract}
The non-perturbative implications of BRST quantization are examined for
$\theta$ vacua and related issues. Strong CP violation is shown to be absent
for QCD in the BRST formalism. Previous evidence for CP violation is
reexamined, and much of it is found to be inconclusive.
It is proposed that some lattice
calculations be redone to clarify the situation. For the U(1) problem,
difficulties are encountered for a conventional solution within the
BRST framework. We also find problems with previous instanton and
non-instanton approaches.
\end{abstract}
 \narrowtext

{\bf 1. Introduction}
\vskip .5cm

BRST invariance \cite{brst} has proven to be a powerful tool to establish the
renormalizability of the standard model \cite{sm} and to elucidate its formal
structure \cite{kugo}. It has been subsequently extended to encompass general
gauge theories \cite{fravil} and string theories \cite{string}.

The purpose of this paper is to examine the non-perturbative implications
of BRST quantization for QCD. For QCD in the BRST formalism, specific
conclusions can be reached regarding
the nature of $\theta$ vacua, strong CP violation and the
U(1) problem. Taken at face value, there appears to be some evidence
against these results. Upon closer scrutiny, however, this evidence is found
to be inconclusive. As a result, we also find that the U(1) problem is still
a problem. Previous instanton and non-instanton approaches have problems.

The organization of the paper is as follows.
In section 2, we discuss the strong CP problem. We show that strong CP
violation is absent with BRST quantization. Some other corollaries are also
derived. In section 3, we discuss the U(1) problem. Difficulties are
encountered in finding a conventional solution within the BRST framework.
In section 4, previous evidence for CP violation is reanalyzed and much of it
is found to be inconclusive. In particular, existing instanton calculations
in singular gauges are shown to violate divergence identities,
and chiral perturbation theory is found to break down away from $\theta =0$.
Problems with previous instanton and non-instanton approaches to the U(1)
problem are also noted. In section 5, the validity of the BRST quantization
itself is discussed, and its possible inequivalence with other quantization
schemes is considered. Specific suggestions for new lattice calculations are
suggested, that may shed light on the presently discussed issues.

In the discussion, we will occasionally
need to switch between Minkowski space and Euclidean
space. The notation $A_{\mu} B^{\mu}$
will be used for Minkowski space and $A_{\mu}B_{\mu}$ for Euclidean space,
to distinguish between the two when necessary.

\vskip 1cm
{\bf 2. Strong CP Violation}
\vskip .5cm

The discovery of instantons \cite{bel} has led to the problem of strong
CP violation.
Generally one has tried to eliminate the problem by some relaxation mechanism
for the vacuum angle. If the relaxation mechanism is due to weak interactions
\cite{pec}, one has axions \cite{wei}. However, empirical evidence does not
favor this possibility so far. There have also been attempts to find a
mechanism within QCD itself \cite{sam,kni}. However, decisive results are
difficult
to obtain, due to the intractability of QCD at long distances.

In this section, we show that such an explicit mechanism may be unnecessary.
For QCD in the BRST formalism, there is no strong CP violation in the first
place.

We recall that the topological charge density
$\Xi=(g^2/32\pi^2)F^a_{\mu\nu}\tilde F^{\mu\nu a}$ is a total divergence
$\Xi =\partial^{\mu} K_{\mu}$, where
\be
K_{\mu} =\frac {g^2}{32\pi^2} \epsilon_{\mu\nu\rho \sigma}
(A^{\nu a} F^{\rho \sigma a} -\frac g3 f^{abc} A^{\nu a} A^{\rho b}
A^{\sigma c} )
\label{1}
\ee
is the Loos-Chern-Simons current. It follows that the QCD Hamiltonian for
vacuum angle $\theta$ is related to that for $\theta =0$
by a unitary transformation
$H(\theta ) = e^{i\theta {\bf X}} H(0) e^{-i\theta {\bf X}}$ with
${\bf X} = \int d^3x K_0 $. ${\bf X}$ is invariant under infinitesimal gauge
transformations, so it is invariant under BRST transformations
\footnote{Note that here ${\bf X}$ is single-valued over the physical subspace,
while in the canonical formalism (see below) ${\bf X}$ is multi-valued.}.
Hence, if $|0>$ is the physical ground state of $H(0)$, $i.e.$
\be
H(0) |0> = 0 \qquad\qquad Q_{\rm BRST} |0> =0
\label{2}
\ee
then, $e^{i\theta {\bf X}} |0>$ is the physical ground state of $H(\theta )$,
\be
H(\theta ) e^{i\theta {\bf X}} |0> = 0 \qquad\qquad
Q_{\rm BRST} e^{i\theta {\bf X}} |0> =0
\label{3}
\ee
with the same energy.
In general, $H(\theta )$ is physically equivalent to
$H(0)$, and there is no CP violation in particular.

The point is that there seems to be no constraints corresponding to
topologically non-trivial ("large") gauge transformations. The BRST Hamiltonian
is not invariant under ordinary "large" gauge transformations, and there
are no "large" BRST transformations, since BRST transformations are global
(rigid)
\footnote{On compact spacetime
manifolds, "twisted" gauge transformations exist \cite{baal} which do not
commute with covariant gauge fixing. However, such transformations introduce
external fluxes into the system and will not be considered here.}.

The $\theta$-independence of the
vacuum energy density ${\cal E}$ in BRST quantization implies $<\theta | \Xi
|\theta > =-\partial{\cal E} /\partial \theta = 0$. This result also follows
from translation invariance \cite{yam}. If the vacuum state is normalizable and
translation invariant
\be
<\theta | K_{\mu} (x) |\theta > = &&
<\theta | e^{iP\cdot x} \, K_{\mu} (0) \, e^{-iP\cdot x} |\theta >\nonumber\\
=&& <\theta | K_{\mu} (0) |\theta >
\label{x3}
\ee
is constant, where $P_{\mu}$ is the energy-momentum operator. Hence,
$<\theta |\Xi |\theta > = \partial^{\mu} <\theta | K_{\mu} |\theta > =0$.

The above
derivation fails to go through in the canonical formalism, where the vacuum
state is not normalizable, so that $<\theta | K_{\mu} |\theta >$ may be
ambiguous like $<p | x |p >$ in quantum mechanics. However, there is no
difficulty in covariant gauge, where the vacuum state is normalizable.

It follows that the topological susceptibility at zero momentum
\be
\chi (\theta ) =&& \frac{\partial <\theta | \Xi |\theta >}{\partial\theta}
\nonumber\\=&&-i\int d^4x \partial^{\mu} \partial^{\nu}
<\theta | T^* K_{\mu} (x) K_{\nu} (0) |\theta > _{\rm con}
\label{16}
\ee
is also zero.

In the above, we have used the fact that derivatives come outside of the
time-ordered product in a path integral formulation. This is necessary for
the Fujikawa-Vergeles analysis \cite{fuj} to be compatible with the Ward
identities for the gauge-variant flavor-singlet axial current
$j^{\rm sym}_{\mu 5}$ \cite{bar}. For an operator ${\cal O}$
with chirality $\alpha$, the former gives
\be
&&\int [dA] [dq] [d\overline{q} ]\,e^{i S[A, q, \overline q ] }
\nonumber\\
&&( \partial^{\mu} j_{\mu 5}^{\rm inv} (x) \,{\cal O} (0)
+ \alpha \,\delta^4 (x) \,{\cal O} (0) -2N_f \Xi (x) \,{\cal O} (0) ) = 0
\label{17}
\ee
in the chiral limit, where
$j_{\mu 5}^{\rm inv} = j_{\mu 5}^{\rm sym} + 2N_f K_{\mu}$
is the gauge-invariant axial current and $N_f$ is the number of flavors.
On the other hand, the latter gives
\be
\partial^{\mu} <\theta | T^* j_{\mu 5}^{\rm sym} (x) {\cal O} (0) |\theta > =
-\alpha \,\delta^4 (x)\,<\theta |{\cal O} (0) |\theta >
\label{18}
\ee
Hence, (\ref{17}) should be interpreted as
\be
&&\partial^{\mu} <\theta |T^* j_{\mu 5}^{\rm inv} (x) \,{\cal O} (0) |\theta >
+ \alpha\, \delta^4 (x)\, <\theta | {\cal O} (0) |\theta >  \nonumber\\
&&= 2N_f \partial^{\mu} <\theta | T^* K_{\mu} (x) \,{\cal O} (0) |\theta >
\label{19}
\ee

Another important check is that (\ref{16}) is invariant under additive
renormalization as it affects only the contact term in
$<\theta | T^* K_{\mu} (x) K_{\nu} (0) |\theta >$, which is annihilated by
$\int d^4x \partial^{\mu} \partial^{\nu}$. With this in mind, we note that
at finite momentum, the correlation function
\be
-i\int d^4x e^{iq\cdot x}\partial^{\mu} \partial^{\nu}
<\theta | T^* K_{\mu} (x) K_{\nu} (0) |\theta > _{\rm con}
\label{a1}
\ee
obeys a twice-subtracted dispersion
relation. Therefore, the vanishing of the topological susceptibility (\ref{16})
provides no useful information on the spectrum, apart from the prohibition of
massless vector ghosts.

Another corollary is obtained by taking the vacuum expectation value of the
anomaly,
\be
\partial^{\mu} <j_{\mu 5}^{\rm inv} >_{\theta} =
 2m\, < \overline{q} i\gamma_5 q >_{\theta} +2N_f
<  \Xi  >_{\theta}
\label{14}
\ee
where we have taken real and equal quark masses for simplicity.
The vacuum expectation value  of a gauge-invariant operator should be
well defined and translation invariant in any gauge, so the left-hand side is
zero. With $<\Xi >_{\theta} =0$, we have
\be
<\theta |\overline{q} i\gamma_5 q |\theta > = 0 \qquad\qquad m\neq 0
\label{x14}
\ee

\vskip 1cm
{\bf 3. The U(1) Problem}
\vskip .5cm

The axial U(1) symmetry of the QCD action is not apparent in nature, although
a non-vanishing quark condensate suggests the existence of a ninth
Nambu-Goldstone mode in the chiral limit \cite{gla}. This is the well-known
U(1) problem. The absence of such a mode in the hadronic spectrum is believed
to be related to the chiral anomaly \cite{adl}. 't Hooft \cite{hoo} has
suggested that instantons may provide a solution to the problem through the
anomaly (see, however \cite{cre}). Witten \cite{wit} has
proposed that the problem can be solved in the large $N_c$ limit, where the
anomaly can be treated as a perturbation. Witten's proposal was later
interpreted by Veneziano \cite{ven} in terms of vector ghosts.

Note, however, that the large $N_c$ analysis for QCD rests on
the assumption that the
topological susceptibility is non-zero in Yang-Mills theory, which is
incompatible with BRST quantization as we have seen. In particular,
Veneziano's analysis involves
the assumption that translation invariance is broken for
the one-point function of $K_{\mu}$ but not for the two-point function.

A non-zero $< \Xi  >_{\theta}$ also tends to run counter
to large $N_c$ arguments. The Eguchi-Kawai reduction \cite{egu}
implies that translation invariance should be maintained on the lattice for
large $N_c$. Also, the master field $A_{\theta}^M$
(in the weaker sense of \cite{haa}) for large $N_c$ Yang-Mills theory is
expected
to be translation invariant \cite{wit5}. It would then follow that
\be
<\theta | \Xi |\theta  > = \Xi [A_{\theta}^M ] =\partial^{\mu} (K_{\mu}
[A_{\theta}^M ] ) = 0
\label{22}
\ee

Another piece of evidence is the vacuum energy density ${\cal E}$.
$N_c$ counting arguments give ${\cal E} = N_c^2
\,F(\theta/N_c)$ to leading order, where $F$ is some function. For this to be
periodic in $\theta$ with a period independent of $N_c$, $F$ must be a
constant. Hence $<\Xi >_{\theta} =-\partial{\cal E} /\partial \theta =0$ to
leading order.

To avoid this conclusion, Witten \cite{wit} has previously argued that
$F$ is a multi-valued function. In as much as the original premise of
$N_c$ counting relies on the fact that amplitudes are given by an infinite sum
of Feynman diagrams which is necessarily single-valued, one is at a
dilemma. On the other hand,
Leutwyler and Smilga \cite{leut} have argued that the period of $\theta$
is indeed $2\pi N_c$ for Yang-Mills theory. However, such a result is contrary
to the canonical formalism (see below) as well as the BRST formalism.

We therefore reexamine the U(1) problem within the BRST formalism. The absence
of a physical massless U(1) boson for massive quarks gives
\be
0 = && \int d^4x \,\partial^{\mu}\,
<\theta | T^* j_{\mu 5}^{\rm inv} (x) \overline{q} i\gamma_5 q (0) |\theta >
\nonumber\\
=&&-2i <\theta |\overline{q} q |\theta > \nonumber\\&&
+ 2m \int d^4x <\theta | T^* \overline{q}i\gamma_5 q(x)
\overline{q} i\gamma_5 q (0) |\theta >\nonumber\\
&&+2N_f \int d^4x \partial^{\mu}
<\theta | T^* K_{\mu} (x) \overline{q} i\gamma_5 q (0) |\theta >
\label{30}
\ee
where we have used the massive analogue of (\ref{19}).
On the other hand, (\ref{x14}) gives
\be
&&\int\, d^4x\, \partial^{\mu}
<\theta | T^* K_{\mu} (x) \overline{q} i\gamma_5 q (0) |\theta > \nonumber\\
&&=-i\frac{\partial}{\partial\theta} <\theta | \overline{q} i\gamma_5 q |\theta
> = 0
\label{31}
\ee
Hence
\be
&&-2i <\theta |\overline{q} q |\theta > \nonumber\\
&&+ 2m \int d^4x
<\theta | T^* \overline{q}i\gamma_5 q(x)
\overline{q} i\gamma_5 q (0) |\theta > = 0
\label{32}
\ee

However, this is just the type of equation we would obtain for the flavor
non-singlet current
\be
0=&&\int d^4x\, \partial^{\mu} <\theta | T^*
{\overline q} \gamma_{\mu}\gamma_5 \frac{\lambda^a}2 q (x) \overline{q}
i\gamma_5 \lambda^b q (0) |\theta >\nonumber\\
=&&-\frac i2 <\theta | \overline q [\lambda^a,\lambda^b ]_+ q |\theta >
\nonumber\\
&&+m \int d^4x <\theta | T^* \overline{q}i\gamma_5 \lambda^a q(x)
\overline{q} i\gamma_5 \lambda^b q (0) |\theta >
\label{33}
\ee
Therefore it appears that the same conventional assumptions which lead to
$m_{\pi}^2 ={\cal O} (m)$ for the latter imply that the mass squared of
the flavor-singlet boson is also ${\cal O} (m)$
\footnote{
Since the isosinglet correlator obeys a twice-subtracted dispersion relation,
it may be that one of the subtraction constant diverges in the chiral limit,
thereby solving the U(1) problem. This behavior, however, is at odds with
conventional QCD perturbation theory. Also, as the pionic thresholds move to
zero in the chiral limit, they may pile up in the
flavor-singlet correlator causing a $1/m$ singularity, and thereby solving the
U(1) problem. The new problem, however, is then why this does not happen for
the flavor non-singlet correlator.}.

In general, (\ref{14}) implies that (\ref{32}) is proportional to $\chi$, so
our result is consistent with a previous observation \cite{svz} of a
trade-off between the solution of the strong CP problem and the solution of
the U(1) problem. Samuel \cite{sam} has
suggested that a simultaneous resolution may be possible, but this does not
appear to be the case.

The point may be seen in another way. The difference between the flavor-singlet
and non-singlet correlator amounts to the OZI violating contribution
\be
<<{\rm Tr} ( i\gamma_5 \, S_A (x,x) )\,\,
{\rm Tr} ( i\gamma_5 \, S_A (0,0) )>>_{\theta }
\label{b1}
\ee
where $S_A (x, y)$ denotes the quark propagator in a fixed
gluon-background $A$, the trace is over internal indices, and the
averaging is over the gluon field only
\footnote{We have suppressed the quark determinant, for convenience.}. Writing
\be
S_A (x,y) =&&\sum \frac{{\psi_n (x)}{\psi_n^{\dagger} (y)}}{i\lambda_n -m}
\nonumber\\
=&&-\frac 1m \sum_{\lambda_n =0} \psi_n (x)\psi_n^{\dagger} (y)
+ \overline{S}_A (x,y)
\label{b2}
\ee
where $\overline{S}_A$ is the non-zero mode part of $S_A$, the index theorem
yields
\be
&&\frac 1V\int d^4x\,d^4y
<<{\rm Tr} ( i\gamma_5 \, \overline{S}_A (x,x) )\,\,
{\rm Tr} ( i\gamma_5 \, \overline{S}_A (y,y) )>>_{\theta } \nonumber\\=&&
-\frac{N_f^2}{Vm^2} <<\nu^2>>_{\theta} \nonumber\\&&+
\frac{2iN_f}{Vm}\int d^4y
<< \nu\, {\rm Tr} ( i\gamma_5 \, \overline{S}_A (y,y) )>>_{\theta }\nonumber\\
&&+\frac 1V \int d^4x\, d^4y
 <<{\rm Tr} ( i\gamma_5 \, {S}_A (x,x) )\,\,
{\rm Tr} ( i\gamma_5 \, {S}_A (y,y) )>>_{\theta }\nonumber\\
\label{b3}
\ee
where $V$ is the spacetime volume and
$\nu = \int_V \, d^4x \, \Xi$ is the topological charge.
Hence, the vanishing of the topological susceptibility (\ref{16})
and (\ref{31}) imply the absence of
$1/m^2$ and $1/m$ singularities which could have made a difference.

\vskip 1cm
{\bf 4. Reexamination of Previous Arguments}
\vskip .5cm

There are five reasons  why CP violation was previously assumed to be
present.

The first reason is instantons. Instanton
calculations \cite{cal,dia} give the vacuum energy density as a non-trivial
function of $\theta$. However, the results are not
self-consistent, since they are based on
't Hooft 's calculation \cite{hoo} which employs BRST quantization.

A possible source of this inconsistency is that the
calculations in \cite{cal,dia}
adopt a singular gauge to evaluate the instanton-antiinstanton
interaction
\footnote{To our knowledge, no one has evaluated
the instanton-antiinstanton interaction in the original (regular)
gauge of Ref. \cite{bel}, owing to difficulties with the long range tail.}.
To see this, let $D$ be a domain
in Euclidean space bounded by two surfaces, one $S_1$ surrounding the
singularity, and another $S_2$ at a large distance away, so that
\be
\int_D \, d^4x\, \Xi (x) =
\oint_{S_2}\, d^3 S_{\mu} \,K_{\mu} - \oint_{S_1}\, d^3 S_{\mu} \,K_{\mu}
\label{4}
\ee
For an instanton in singular gauge, the fields fall off rapidly at infinity so
that the integral over $S_2$ vanishes as $S_2$ is sent off to infinity.
Shrinking $S_1$ then gives
\be
{\rm lim}\, \oint_{S_1}\, d^3S_{\mu} \,K_{\mu} = -1
\label{5}
\ee
However, this means that $\partial_{\mu} K_{\mu} = \Xi -\delta^4$, thus
violating the basic identity.

The second reason is canonical quantization. The Hamiltonian still obeys
$H(\theta ) = e^{i\theta {\bf X}} H(0) e^{-i\theta {\bf X}}$, but the
subsidiary condition is altered to
\be
{\bf U}[\Omega ] \,\,|{\rm phys} > = |{\rm phys}>
\label{6}
\ee
where ${\bf U} [\Omega ]$ is the unitary operator implementing the gauge
transformation $\Omega$ (unitarized form of Gauss law).
If $\Omega$ is topologically non-trivial,
$e^{i\theta {\bf X}}$ does not commute with the constraint (\ref{6}) unless
$\theta$ is an integer multiple of $2\pi$. Hence, Hamiltonians with $\theta
\neq 0$ were regarded as physically inequivalent. However, the argument
applies independently of the presence of quarks, whereas physics should be
independent of $\theta$ if massless quarks are present
\footnote{
The expectation value of operators with net chirality such as the quark
condensate does exhibit $\theta$ dependence. However, this does not matter
for the physics.}, so it is not generally reliable.

The third reason is an effective Lagrangian calculation \cite{leut}, in which
the vacuum energy density is given by
\be
{\cal E}_0 = -\Sigma \bigg( m_u^2 +m_d^2 + 2 m_um_d \,\,{\rm cos} \theta
\bigg)^{1/2}
\label{7}
\ee
where $m_u$ and $m_d$ are the up and down quark masses, and $\Sigma = -
<0 |\overline{u} u |0 >$ in the chiral limit.

However, the effective Lagrangian
approach relies on chiral perturbation theory, which is not valid for
$\theta \neq 0$ if BRST quantization is valid. For example, the
Fujikawa-Vergeles analysis \cite{fuj}  gives
\be
&&\lim_{{\hat m}\to0^+}
<\theta, \hat{m}\,e^{-i\theta \gamma_5/N_f} |
\overline{q} q |\theta, \hat{m}\, e^{-i\theta \gamma_5/N_f} > \nonumber\\
&&= -N_f\Sigma {\rm cos} \,\,\frac{\theta}{N_f}
\nonumber\\
&&\lim_{{\hat m}\to0^+}
<\theta, \hat{m}\, e^{-i\theta \gamma_5/N_f} |
\overline{q}i\gamma_5 q |\theta, \hat{m}\, e^{-i\theta \gamma_5/N_f} >
\nonumber\\
&& = +N_f\Sigma \,\,{\rm sin} \frac{\theta}{N_f}
\label{12}
\ee
where we have written out the quark mass matrix  explicitly.
According to conventional wisdom $\Sigma \neq 0$, so comparison with
(\ref{x14})
implies
\be
&&\lim_{{\hat m}\to0^+}
<\theta, \hat{m}\, e^{-i\theta \gamma_5/N_f} |
\overline{q}i\gamma_5 q |\theta, \hat{m}\, e^{-i\theta \gamma_5/N_f} >
\nonumber\\
&&\neq \lim_{ m\to0}
<\theta, m| \overline{q}i\gamma_5 q |\theta, m >
\label{13}
\ee
in general. Hence, chiral perturbation theory breaks down for $\theta
\neq 0$.

A weaker but gauge-independent argument may also be given.
 Eq. (\ref{14}) implies that $<\Xi >_{\theta}$ must be ${\cal O} (m )$
for chiral perturbation theory to hold.
On the other hand, the $m$-dependence of $<\Xi >_{\theta}$ enters
only through the determinant obtained after integrating out the quarks, $i.e.$
\be
<\Xi >_{\theta} = \frac{<<\Xi \bigg({\rm det} (\gamma_{\mu} \nabla_{\mu}
-m)\bigg)^{N_f} >>_{\theta}}{<<\bigg({\rm det} (\gamma_{\mu} \nabla_{\mu}
-m)\bigg)^{N_f} >>_{\theta}}
\label{15}
\ee
where ${\rm det} (\gamma_{\mu} \nabla_{\mu} -m)$ is the determinant for one
flavor and $<<\,\,\,>>_{\theta}$ denotes the average over gluon fields only.
However, it is seen that (\ref{15}) is not generally ${\cal O} (m)$. Indeed,
it is ${\cal O} (1)$ in the quenched approximation $N_f=0$.

The fourth reason is the topological susceptibility. The BRST formalism gives
$\chi =0$ as we have seen.
On the other hand, lattice calculations give a non-zero topological
susceptibility in Yang-Mills theory \cite{lat} and QCD \cite{lat1} for zero
vacuum angle. However, almost all of them use periodic boundary conditions
which is problematic. In the continuum limit, this would imply a sharp (zero)
topological charge $\nu$ independently of $\theta$, and
hence zero topological susceptibility.

There are two reasons why the lattice calculations yield a non-zero $\chi$.
Most of the lattice calculations \cite{lat,lat1}
employ a lattice topological density
$\Xi_L$ \cite{lus} in a conventional Monte-Carlo simulation to compute
$<0|T^* \Xi_L (x) \Xi_L (0) |0>$. However, such a calculation can differ from
the correct result by contact terms. Specifically, in the continuum limit
\be
&&-\partial^{\mu}\partial^{\nu} <0| T^* K_{\mu} (x) K_{\nu} (0) |0> \nonumber\\
&&=<0|T^* \Xi_L (x) \Xi_L (0) |0 > -\alpha_* \, \delta^4 (x) -\beta_*\,
\partial^2
\, \delta^4 (x) - ...
\label{20}
\ee
A vanishing $\chi (0)$ implies that
\be
\chi_L (0) = \int_V d^4x <0|T^* \Xi_L (x) \Xi_L (0) |0> =\alpha_*
\label{21}
\ee
We also note that unlike $\chi (0)$, the lattice result (\ref{21}) is scheme
dependent and can be renormalized to an arbitrary value.

To remedy this problem, some calculations have used cooling procedures
to smooth out the lattice gauge configurations, washing them
out of their ultraviolet content. After several cooling sweeps, bare
topological susceptibilities were reported.
However, the gauge field configurations with non-zero lattice topological
charge must correspond to singular gauge configurations in the continuum limit
\cite{jan}. The latter carry $\Xi_L
\neq \partial_{\mu} K_{\mu}$ in the continuum, but this is
unacceptable since the operator identity $\Xi = \partial^{\mu} K_{\mu}$
requires all configurations in the path integral to obey the same identity
\footnote{
Note that singular gauge transformations are also excluded in the canonical
formalism (\ref{6}).}.
We note that the cooled lattice calculations \cite{lat,lat1},
when rid of the specific gauge configurations for which
$\nu_L=\int_V d^4x <\Xi_L > \neq 0$ give
$\chi_L =0$ \footnote{Of course, the averaging over all gauge
configurations give necessarily zero for the total $\nu_L$, since the vacuum
for $\theta =0$ is parity even.}.

The fifth reason is 1+1 dimensional models, the simplest example being free
Maxwell theory. The analog of $\Xi$ is then the electric field which can be a
non-zero constant. However, this is due to the fact that the analog of
$K_{\mu}$ is ill-defined in 1+1 dimensions. The infrared divergence $\int \,
d^2k/k^2$ which makes $K_{\mu}$ ill-defined is peculiar to $1+1$ dimensions,
and it is not clear whether the example is relevant to 3+1 dimensions.

To summarize, the current evidence for strong CP violation appears to be
inconclusive.
As far as the U(1) problem is concerned, 't Hooft's calculation relies on BRST
quantization. Since the existing instanton calculations of the vacuum energy
are at odds with the BRST result, instantons cannot be said to solve the
U(1) problem without further amendment. Given that Veneziano's analysis
assumes translation invariance for the two-point function $K_{\mu}$ but not for
the one-point function, we conclude that the U(1) problem is still a problem.

\vskip 1cm
{\bf 4. Conclusions}
\vskip .5cm

We have assumed the validity of BRST quantization throughout. However,
it is well-known that the Faddeev-Popov prescription suffers from the Gribov
ambiguity at the non-perturbative level, so there is reason for concern about
such an assumption. However, the BRST Lagrangian for $Q_{\rm BRST} =0$ is
equivalent to QCD, and gauge independence follows from
the Fradkin-Vilkovisky theorem \cite{fravil}. Hence, BRST quantization should
represent a legitimate quantization scheme for QCD. Indeed, 't Hooft's
instanton calculations require the validity of BRST quantization at
the non-perturbative level.

It is also important that the BRST formalism holds outside of QCD.
Indeed, if it were not the case, one may ask whether the vacuum
expectation value of the Higgs field (another gauge-variant object) could break
translation invariance, and if so, what happens to the W and Z mass.

The simplest possibility is that BRST quantization may be inequivalent to
other quantization schemes at the non-perturbative level.
If we turn to other models, it is reasonably certain that the $2+1$ dimensional
Polyakov model \cite{pol} has no $\theta$ dependence in the physics, since the
analog of $K_{\mu}$ is gauge invariant, and $e^{i\theta {\bf X}}$ commutes with
the constraint in both the BRST formalism and the canonical formalism.
On the other hand, canonical reasoning implies $\theta$ dependence in the
monopole sector of the Yang-Mills-Higgs system \cite{wit2}, so this is a case
where a difference is expected with the BRST formalism. As we did note,
however,
the canonical formalism is inapplicable in the presence of quarks
\footnote{A modified canonical formalism incorporating the anomaly has been
discussed by Manton \cite{man}, for the massless Schwinger model on a circle.}.

Given the above, it is desirable to perform new lattice calculations of the
topological susceptibility to ascertain whether lattice quantization
is equivalent (or not) with the
BRST quantization for Yang-Mills theory and QCD. For
reasons stated above, the calculations should employ  free boundary
conditions, and a large enough lattice to avoid edge effects. The lattice
configurations should be explicitly checked against contaminations by singular
gauge configurations (dislocations as well as divergence violating ones).
The results should give us a better idea whether the problem massive QCD really
has to face is strong CP violation or the U(1) problem.

\vglue 0.6cm
{\bf \noindent  Acknowledgements \hfil}
\vglue 0.4cm
This work was supported in part  by the US DOE grant DE-FG-88ER40388.

\vskip 1cm
\setlength{\baselineskip}{15pt}


\begin{thebibliography}{50}


\bibitem{brst}
C. Becchi, A. Rouet and R. Stora, Comm. Math. Phys. {\bf 42} (1975) 127;
Ann. Phys. {\bf 98} (1976) 287; I.V. Tyutin, Lebedev Preprint FIAN39 (1975)
unpublished.



\bibitem{sm}
G.'t Hooft, Nucl. Phys. {\bf B33} (1971) 173; Nucl. Phys. {\bf B35} (1971) 167;
G. 't Hooft and M. Veltman, Nucl. Phys. {\bf B44} (1972) 189; B.W. Lee and J.
Zinn-Justin, Phys. Rev. {\bf D5} (1972) 3121; {\bf D7} (1973) 1049; B.W. Lee,
Phys. Rev. {\bf D9} (1974) 938.


\bibitem{kugo}
T. Kugo and I. Ojima, Prog. Theor. Phys. Suppl. {\bf 66} (1979) 1.


\bibitem{fravil}
E.S. Fradkin and G.A. Vilkovisky, Phys. Lett. {\bf B55} (1975) 224;
I.A. Batalin and G.A. Vilkovisky, Phys. Lett. {\bf B69} (1977) 309.


\bibitem{string}
M. Kato and K. Ogawa, Nucl. Phys. {\bf B212} (1983) 443;
S. Hwang, Phys. Rev. {\bf D28} (1983) 2614;
W. Siegel, Phys. Lett. {\bf B149} (1984) 157, 162.


\bibitem{bel}
A.A. Belavin, A.M. Polyakov, A.S. Schwarz and Yu. S. Tyupkin, Phys. Lett. {\bf
B59} (1975) 85.


\bibitem{pec}
R. Peccei and H. Quinn, Phys. Rev. Lett. {\bf 38} (1977) 1440.

\bibitem{wei}
S. Weinberg, Phys. Rev. Lett. {\bf 40} (1978) 223;
F. Wilczek, Phys. Rev. Lett. {\bf 40} (1978) 279.

\bibitem{sam}
S. Samuel, Mod. Phys. Lett. {\bf A7} (1992) 2007.

\bibitem{kni}
P. Minkowski, Phys. Lett. {\bf B76} (1978) 439;
V.C. Knizhnik and A.M. Morozov, JETP Lett. {\bf 39} (1984) 241;
G. Schierholz, DESY preprints 94-031 and 94-229 (1994).



\bibitem{baal}
G. 't Hooft, Comm. Math. Phys. {\bf 81} (1981) 267;
P. Van Baal, Comm. Math. Phys. {\bf 85} (1982) 529; Comm.
Math. Phys. {\bf 94} (1984) 397;
P.J. Braam and P. Van Baal, Comm. Math. Phys. {\bf 122} (1989) 267;



\bibitem{yam}
H. Yamagishi, Prog. Theor. Phys. {\bf 87} (1992) 785.


\bibitem{fuj}
K. Fujikawa, Phys. Rev. Lett. {\bf 42} (1979) 1195;  {\bf 44} (1980) 1733;
S.N. Vergeles, as quoted in A.A. Migdal, Phys. Lett. {\bf 81B} (1979) 37.

\bibitem{bar}
W.A. Bardeen, Nucl. Phys. {\bf B75} (1974) 2342.


\bibitem{gla}
S.L. Glashow, in Hadrons and their Interactions, edited by A. Zichichi,
Academic, N.Y. 1968;
M. Gell-Mann, in Proceedings of the Third Hawaii Topical Conference, edited by
S.F. Huan, Western Periodicals, North Hollywood, California, 1970.

\bibitem{adl}
S.L. Adler, Phys. Rev. {\bf 177} (1969) 2426;
J.S. Bell and R. Jackiw, Nuov. Cim. {\bf 60A} (1969) 47.

\bibitem{hoo}
G. t' Hooft, Phys. Rev. Lett. {\bf 37} (1976) 8; Phys. Rev. {\bf D14} (1976)
3432; (E) {\bf D18} (1978) 2199; Phys. Rep. {\bf 142} (1986) 357.

\bibitem{cre}
R.J. Crewther, in Field Theoretical Methods in Particle Physics, edited by W.
Ruhl, Plenum, N.Y. 1980; G.A. Christos, Phys. Rep. {\bf 116} (1984) 251.

\bibitem{wit}
E. Witten, Nucl. Phys. {\bf B156} (1979) 269; Ann. Phys. (NY) {\bf 128} (1980)
363.

\bibitem{ven}
G. Veneziano, Nucl. Phys. {\bf B159} (1974) 213;

\bibitem{egu}
T. Eguchi and H. Kawai, Phys. Rev. Lett. {\bf 48} (1982) 1063.

\bibitem{haa}
O. Haan, Phys. Lett. {\bf B106} (1981) 207.

\bibitem{wit5}
E. Witten, in Recent Developments in Gauge Theories, edited by
G. 't Hooft et al. , Plenum, New-York, 1980.


\bibitem{leut}
H. Leutwyler and A. Smilga, Phys. Rev. {\bf D46} (1992) 5607.


\bibitem{svz}
M.A. Shifman, A.I. Vainshtein and V.I. Zakharov, Nucl. Phys.
{\bf B166} (1980) 439.




\bibitem{cal}
C.G. Callan, R. Dashen and D. Gross, Phys. Rev. {\bf D17} (1978) 2717.

\bibitem{dia}
D.I. Dyakonov and V. Yu. Petrov, Nucl. Phys. {\bf B245} (1984) 259.





\bibitem{lat}
P. Woit, Phys. Rev. Lett. {\bf 51} (1983) 638;
K. Ishikawa, $et$ $al.$, Phys. Lett. {\bf B128} (1983) 309;
M. Teper, Phys. Lett. {\bf B162} (1985) 357;
P. Woit, Nucl. Phys. {\bf B262} (1985) 284;
H. Hoek, M. Teper and J. Waterhouse, Phys. Lett. {\bf 180} (1986) 112;
M. Teper, Phys. Lett. {\bf B171} (1986) 81, 86;
E.M. Ilgenfritz, $et$ $al.$, Nucl. Phys. {\bf B268} (1986) 693;
Y. Arian and P. Woit, Phys. Lett. {\bf B 183} (1987) 341;
A.S. Kronfeld, $et$ $al.$ Nucl. Phys. {\bf B292} (1987) 330;
A.S. Kronfeld, $et$ $al.$ Nucl. Phys. {\bf B305} [FS23] (1988) 661;
M.I. Polikarpov and A.I. Veselov, Nucl. Phys. {\bf B297} (1988) 34;
M. Teper, Phys. Lett. {\bf B202} (1988) 553;
J.C. Vink, Phys. Lett. {\bf B212} (1988) 483;
J. Smith and J.C. Vink, Nucl. Phys. {\bf B202} (1988) 553;
M. Campostrini, $et$ $al.$ Nucl. Phys. {\bf B329} (1990) 683.
B. Alles and M. Gianetti, Phys. Rev. {\bf D44} (1991) 513;
M. Teper, Nucl. Phys. [Suppl.] {\bf B20} (1991) 159;
A. Di Giacomo, Nucl. Phys. [Suppl.] {\bf B23} (1991) 191;
B. Alles, $et$ $al.$, Phys. Rev. {\bf D48} (1993) 2284;
M.C. Chu $et$ $al.$, Phys. Rev. {\bf D49} (1994) 6039;
J. Grandy and R. Gupta, Nucl. Phys. {\bf B34} [Supp.] (1994) 164;
C. Michael and P.S. Spencer, Liverpool Preprint (1995) : LTH 346,
HEP-LAT-9503018.


\bibitem{lat1}
K.M. Bitar, $et$ $al.$, Phys. Rev. {\bf D44} (1991) 2090;
Y. Kuramashi, M. Fukugita, H. Mino, M. Okawa and A. Ukawa,
Phys. Lett. {\bf B313} (1993) 425.

\bibitem{lus}
P. Di Vecchia, K. Fabricius, G.C. Rossi and G. Veneziano, Nucl. Phys.
{\bf B192} (1981) 392; Phys. Lett. {\bf B108} (1982) 323;
M. Luscher, Comm. Math. Phys. {\bf 85} (1982) 29.

\bibitem{jan}
J. Polonyi, Phys. Rev. {\bf D29} (1984) 716.




\bibitem{pol}
A.M. Polyakov, Nucl. Phys. {\bf B120} (1977) 429.

\bibitem{wit2}
E. Witten, Phys. Lett. {\bf B86} (1979) 283


\bibitem{man}
N.S. Manton, Ann. Phys. {\bf 159} (1985) 220.




\end{thebibliography}
\end{document}